\documentclass[a4paper]{jpconf}
\usepackage{graphicx}

\def\vect#1{{\mbox{\boldmath $#1$}}}

\def\ketv#1{\mid \hspace{-0.5mm} #1 \rangle}
\def\brav#1{\langle #1 \hspace{-1mm} \mid}

\begin{document}
\title{Simulation of heavy ion collision \\using a time-dependent density functional theory 
\\including nuclear superfluidity}

\author{Shuichiro EBATA}

\address{Center for Nuclear Study, University of Tokyo, Hongo, Bunkyo, Tokyo 113-0033, Japan}

\ead{ebata@cns.s.u-tokyo.ac.jp}

\begin{abstract}
We carried out a simulation of heavy ion collision using a time-dependent 
density functional theory. 
We call it the canonical-basis time-dependent Hartree-Fock-Bogoliubov 
theory (Cb-TDHFB) which can describe nuclear dynamics 
in three-dimensional coordinate space, treating nuclear pairing correlation. 
We simulate $^{20}$O+$^{20}$O collision using the Cb-TDHFB with a contact-type pairing 
functional, and show the behavior of gap energy which is decreasing and vibrating 
while colliding. 
\end{abstract}

\section{Introduction}
Atomic nucleus is a self-binding finite quantum many-body system whose size is several femtometers
(= fm = 10$^{-15}$m). 
It is composed of two kinds of fermions (protons and neutrons) with spin $1/2$ 
that are called ``nucleons''. 
The number of nucleons in a nucleus is a few hundreds at most. 
There are about 3,000 nuclei observed in experiments, and yet more to be 
discovered.  
In order to study the structure of nuclei, we investigate many excited modes of many nuclei. 
To study nuclear collision microscopically is also very important. 
Nucleus has ``magic numbers'', which are analogous to those in atoms. 
This suggests that the nucleon in the nucleus is freely moving in an attractive 
average potential. 
The nucleons create the mean field by themselves. 

In the time-dependent mean-field theory, 
the heavy-ion collision is expressed as colliding of two mean fields. 
We should investigate the behavior of single-particle states while colliding two fields, 
to expect the structure of nuclei after the heavy-ion collision. 
Nuclear pairing correlation has a significant role to define single-particle states. 
It allows the fractional occupation probability of single-particle states due to 
the pair scattering, which is important to reconstruct a new mean field. 
The time-dependent Hartree-Fock theory (TDHF) is well-known as a useful tool to 
study nuclear dynamics\cite{KD77}, however it can not include effects of pairing correlation. 

There is a theory to treat pairing correlation in nuclear dynamics self-consistently, 
which is called the time-dependent Hartree-Fock-Bogoliubov theory (TDHFB). 
But, there is no study of heavy ion collision using TDHFB with a modern effective 
interaction in three-dimensional space, because the study is hard to prepare 
converged initial state, and it needs the huge computational resource\cite{BL10}. 
We proposed a new TD mean-field theory to study nuclear dynamics treating pairing, 
which is named the canonical-basis TDHFB (Cb-TDHFB)\cite{EN10}. 
The Cb-TDHFB is derived from full TDHFB equations represented in canonical basis 
which diagonalizes density matrix, with Bardeen-Cooper-Schrieffer (BCS)-like 
approximation for pairing functional ($pp$,$hh$-channel). 
We confirmed that linear response calculations of the Cb-TDHFB are a good approximation of 
the quasi-particle random phase approximation (QRPA) which is a small amplitude limit of TDHFB\cite{EN10}.

In the present work, we apply the Cb-TDHFB to heavy ion collision in the three-dimensional Cartesian 
coordinate space using a modern effective interaction (Skyrme: SkM$^{\ast}$) with a contact pairing. 
In the followings, we introduce the Cb-TDHFB equations and the contact pairing functional. 
Then, we show results of $^{20}$O+$^{20}$O which indicate an collision of the gap-energy. 

\section{Formulation and Procedure}
\subsection{Basic equations and contact pairing}
The Cb-TDHFB equations can be derived from the TDHFB equations 
represented in canonical basis with a simple approximation for pairing functional\cite{EN10}. 
The TDHFB equations can be written in terms of the generalized density matrix ${\cal R}(t)$ 
and the generalized Hamiltonian ${\cal H}(t)$\cite{BR86}. 
${\cal R}(t)$ composes the one-body density matrix 
$\rho_{\mu\nu} (t) \equiv \langle \Psi (t) | \hat{c}_{\nu}^{\dag}\hat{c}_{\mu} | \Psi (t) \rangle$ and 
the pairing-tensor $\kappa_{\mu\nu} (t) \equiv \langle \Psi (t) | \hat{c}_{\nu}\hat{c}_{\mu} | \Psi (t) \rangle$: 
$\mu,\nu$ mean arbitrary basis. 
${\cal H}(t)$ has $h(t)$ and $\Delta(t)$ corresponding to 
single-particle Hamiltonian and pair potential, respectively. 
We can express TDHFB state at any time in the canonical (BCS) form as 
\begin{eqnarray}
|\Phi (t) \rangle \equiv 
\prod_{l>0} (u_{l}(t) + v_{l}(t) \hat{c}_{l}^{\dag}(t)\hat{c}_{\bar l}^{\dag}(t) ) | 0 \rangle, 
\label{eq:BCS}
\end{eqnarray}
where $u_{l}(t),v_{l}(t)$ are time-dependent BCS factors and 
$\hat{c}_{l}^{\dag}$ is a creation operator of canonical basis which 
diagonalizes density matrix $\rho (t)$. 
Then, we choose the BCS form of pair potential as
\begin{eqnarray}
\Delta_{l}(t) = - \sum_{k>0} \kappa_{k}(t) \bar{\cal V}_{l\bar{l},k\bar{k}} \ ,
\label{eq:delta}
\end{eqnarray}
where $\kappa_{k}(t) \equiv u_{k}(t)v_{k}(t)$ 
corresponds to the pair tensor $\kappa (t)$ in the canonical-basis 
and $\bar{\cal V}_{l\bar{l},k\bar{k}}$ is the anti-symmetric two-body matrix element. 
The subscripts $\bar{l}$ and $\bar{k}$ mean the pair of canonical basis $l$ and $k$, respectively. 
We can obtain the Cb-TDHFB equations from the canonical form Eq.(\ref{eq:BCS}) of TDHFB, 
and with the pair potential Eq.(\ref{eq:delta}), as follows. 
\begin{eqnarray}
i \hbar \frac{\partial \phi_{l}(t)}{\partial t} &=& (\hat{h}(t) - \eta_{l}(t)) \phi_{l}(t), \ \ i \hbar \frac{\partial \phi_{\bar l}(t)}{\partial t}  = (\hat{h}(t) - \eta_{\bar{l}}(t)) \phi_{\bar{l}}(t),  \nonumber \\
i \hbar \frac{\partial \rho_{l}(t)}{\partial t}  &=& \kappa_{l}(t)\Delta_{l}^{\ast}(t) - \kappa_{l}^{\ast}(t)\Delta_{l}(t),  \nonumber \\
i \hbar \frac{\partial \kappa_{l}(t)}{\partial t}  &=& (\eta_{l}(t) + \eta_{\bar l}(t) ) \kappa_{l}(t) + \Delta_{l}(t) (2\rho_{l}(t) - 1), 
\label{eq:Cb-TDHFB}
\end{eqnarray} 
where $\eta_{l}(t) \equiv \brav{\phi_{l}(t)} \hat{h}(t)\! \ketv{\phi_{l}(t)} + i \hbar \brav{\frac{\partial \phi_{l}}{\partial t}}\! \phi_{l}(t) \rangle$. 
The Cb-TDHFB equations compose the time-evolutions of the canonical basis $\phi_{l}(t), \phi_{\bar l}(t)$, 
the occupation probability $\rho_{l}(t) \equiv |v_{l}(t)|^{2}$ and the pair probability $\kappa_{l}(t)$. 
They conserve orthonormal property of the canonical basis and average particle number. 
When we choose a special gauge condition $\eta_{l}(t) = \varepsilon_{l}(t) = \brav{\phi_{l}(t)}\hat{h}(t) \! \ketv{\phi_{l}}(t)$, 
they conserve average total energy. 

We introduce neutron-neutron and proton-proton BCS pairing. 
The BCS pairing matrix elements ${\cal V}_{l\bar{l},k\bar{k}}^{\tau}$ is written as 
\begin{eqnarray}
 {\cal V}_{l\bar{l},k\bar{k}}^{\tau}=\int\! d\vect{r}_{1}d\vect{r}_{2} \sum_{\sigma_{1},\sigma_{2}} 
 \phi_{l}^{\ast}(\vect{r}_{1},\sigma_{1})\phi_{\bar{l}}^{\ast}(\vect{r}_{2},\sigma_{2})
 \hat{\cal V}^{\tau}(\vect{r}_{1},\sigma_{1};\vect{r}_{2},\sigma_{2}) \nonumber \\
 \times \left[  \phi_{k}(\vect{r}_{1},\sigma_{1})\phi_{\bar{k}}(\vect{r}_{2},\sigma_{2})
 -\phi_{\bar{k}}(\vect{r}_{1},\sigma_{1})\phi_{k}(\vect{r}_{2},\sigma_{2}) \right]. \hspace{-15mm}
\label{eq:Viijj0}
\end{eqnarray}
We introduce the spin-singlet contact interaction to Eq.(\ref{eq:Viijj0}): 
\begin{eqnarray}
\hat{\cal V}^{\tau}(\vect{r}_{1},\sigma_{1};\vect{r}_{2},\sigma_{2}) 
\equiv V_{0}^{\tau} \frac{1-\hat{\vect{\sigma}}_{1}\cdot \hat{\vect{\sigma}}_{2}}{4} \delta(\vect{r}_{1}-\vect{r}_{2}), 
\label{eq:Vint}
\end{eqnarray}
where $\tau$ indicates neutron or proton channel and $V_{0}^{\tau}$ is a strength of pairing\cite{KB90}. 
In this present work, we do not use the time-reversal relation between $l$- and $\bar{l}$-canonical basis. 

\subsection{Procedure and calculation space}
At first, we prepare the initial states (HF or HF+BCS ground states) with 
the wave functions for projectile and target at some impact parameter $b$ 
at distance where they feel only Coulomb interaction. 
Then, we boost the wave functions and translational calculate 
the time-evolution of nuclear densities to obey Eq.(\ref{eq:Cb-TDHFB}). 
In this present work, we use the three-dimensional Cartesian coordinate-space representation for the 
canonical states, $\phi_{l}(\vect{r},\sigma; t) = \brav {\vect{r},\sigma}\! \phi_{l}(t) \rangle$ 
with $\sigma=\pm 1/2$. 
The coordinate space is a box whose x-, y-sides are 30 fm and z-side is 50 fm, 
discretized in the square mesh of $\Delta x = \Delta y = \Delta z = 1.0$ fm. 

\section{Results} 
We simulate the $^{20}$O+$^{20}$O collision with incident energy $E_{\rm int}=40$ MeV 
which is higher than the Coulomb barrier in this system. 
Initial distance between projectile and target is 12 fm with $b=$ 0 fm. 
$^{20}$O ($Z$=8, $N$=12) has superfluidity for neutrons. 
The average gap-energy $\bar{\Delta} \equiv \sum_{l>0} \Delta_{l} / \sum_{l>0}$ is 1.901 MeV. 
The number of canonical basis is 24(1.2 times that of TDHF). 

Figure 1 shows the time evolution of neutron density distributions. 
After the touch ($t=98.66\ {\rm fm\verb|/|c }$: panel (b) in Fig.1), 
the density-distribution vibrates keeping the prolate shape.
This shape of a new mean field corresponds to the shape of $^{40}$S ($Z$=16, $N$=24) 
ground state in intrinsic frame. 
Figure 2 shows the time-evolution of neutron gap-energy for each canonical basis defined in Eq.(\ref{eq:delta}). 
From the touching configuration, the gap energies are decreasing significantly. 
The decrease is similar to the results of old work\cite{CM80}. 
We can see the vibration of gap energies related with nuclear density behavior. 

We simulated the $^{20}$O+$^{20}$O collision using time-dependent mean field theory including 
the effects of nuclear superfluidity, and showed the vibration of gap energies in the collision. 
Currently, we are investigating the detail of gap energy vibration and relation 
with the excited states of $^{40}$S. 
To study the pairing effects for collision phenomena, we should improve our program code 
to parallel one. 
We parallelize the code using MPI for the canonical basis, because 
the canonical basis conserve orthogonality in the Cb-TDHFB which means they do not need 
Gram-Schmidt orthonormalization in the real time-evolution. 
We can expect the parallelization efficiency of the method is good, 
due to the relatively small communication-loss. 

\section*{Acknowledgement} 
This work is supported by HPCI Strategic Program Field 5 "The origin of matter and the universe", 
and for computational resource by the RIKEN Integrated Cluster of Clusters(RICC) and 
by the SR16000 at  YITP in Kyoto University.

\begin{figure}[h]
\begin{center}
\includegraphics[width=140mm, clip]{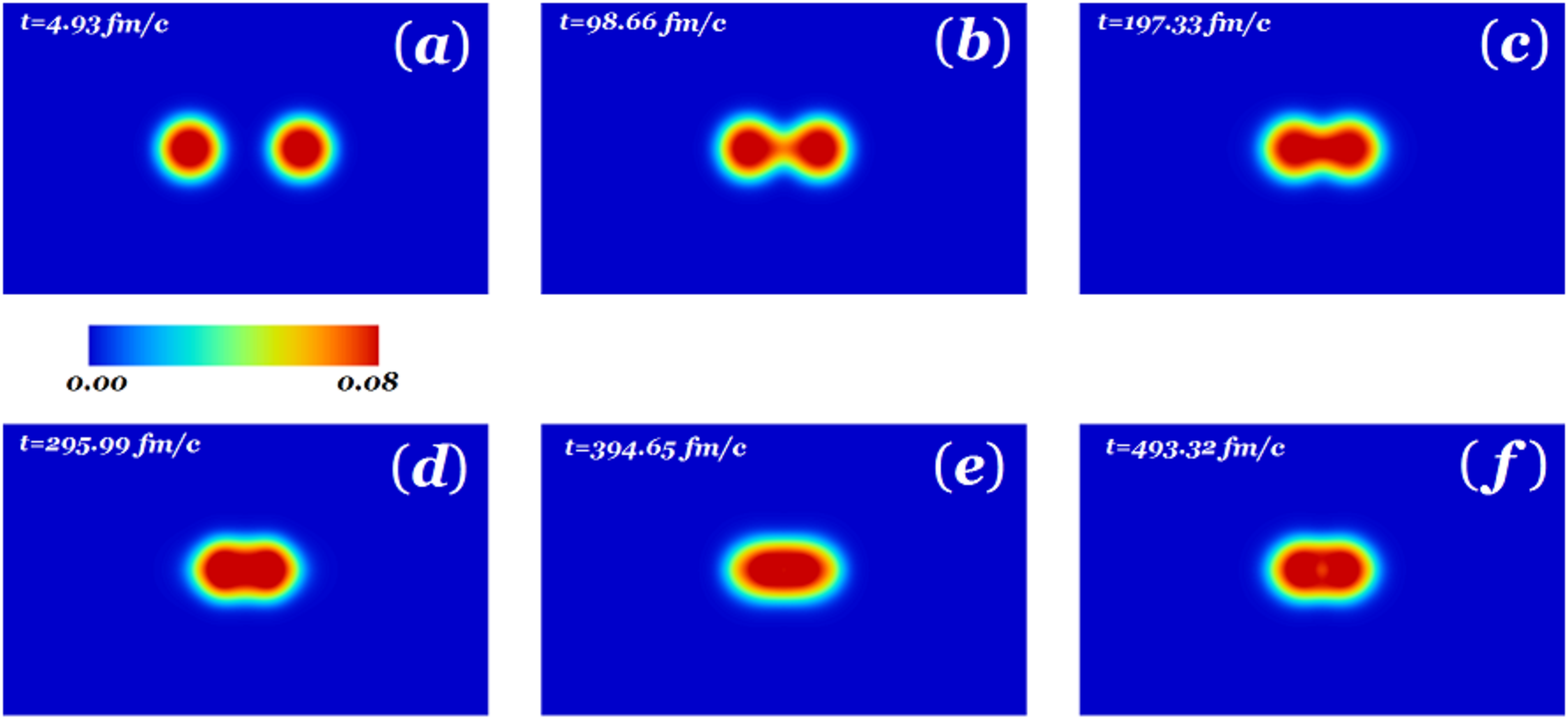}
\caption{Neutron density distributions of XZ-plane in $^{20}$O+$^{20}$O collision 
at $t$=($a$)4.93, ($b$)98.66, ($c$)197.33, ($d$)295.99, ($e$)394.65 and ($f$)493.32 ${\rm fm / c }$.
The horizontal direction corresponds to Z direction.}
\ \\[5mm]
\includegraphics[width=70mm, angle=-90]{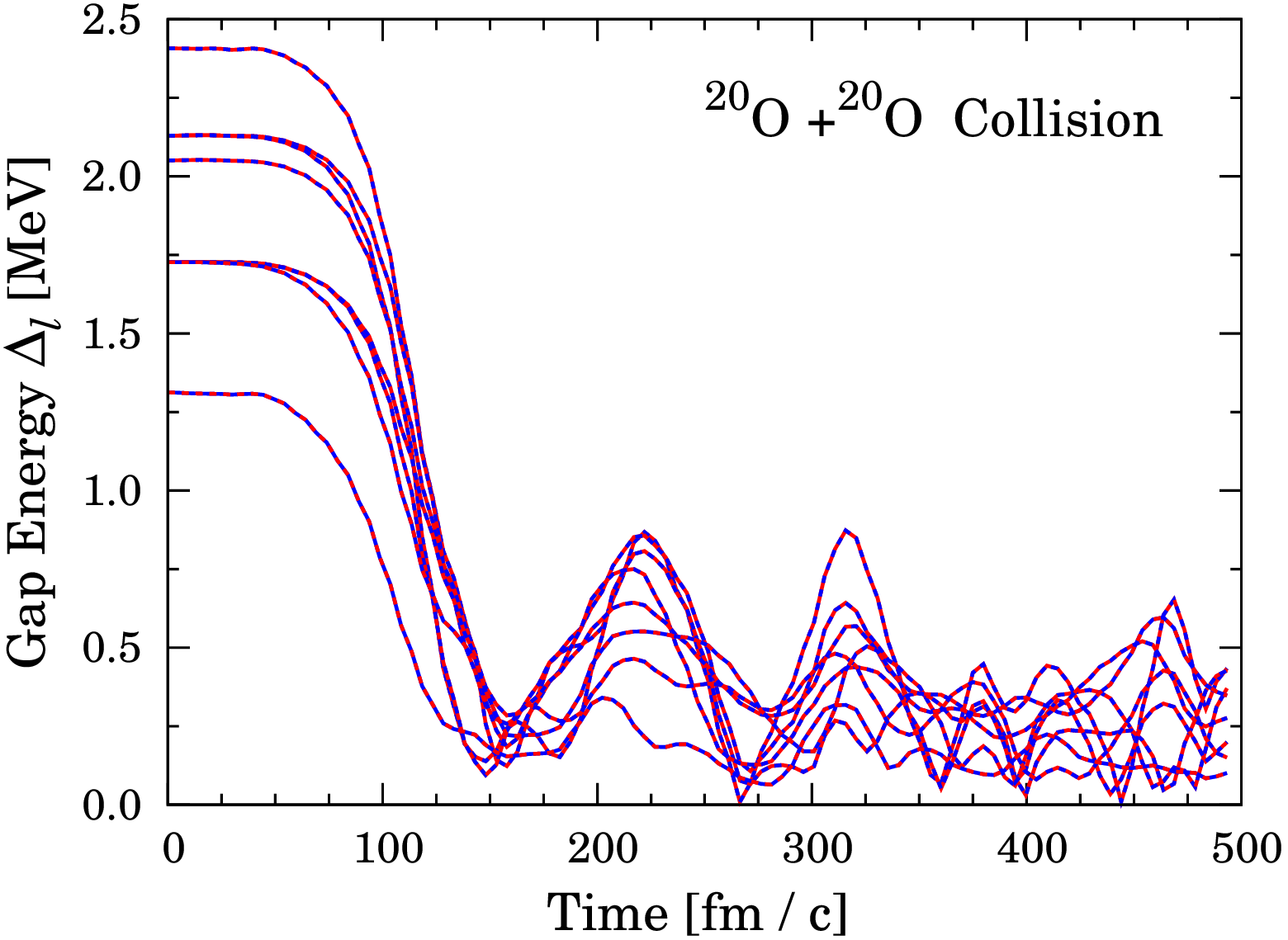}\\[5mm]
\caption{Time-evolution of neutron gap-energy defined in Eq.(\ref{eq:delta}).
Solid lines (red online) and dashed lines (blue online) indicate 
the gap energies of $l$-states in target and projectile nuclei, respectively.}
\end{center}
\end{figure}


\section*{References}
\begin{thebibliography}{9}
\bibitem{KD77} 
S. E. Koonin, K. T. R. Davies, V. Maruhn-Rezwani, H. Feldmeier, S. J. Krieger and J. W. Negele, 
Phys. Rev. C{\bf 15} (1977) 1359.

\bibitem{BL10}
A. Bulgac, Y. L. Luo, P. Magierski, K. J. Roche, I. Stetcu and Y. Yu, 
UNEDF SciDAC Collaboration Research Highlight, 
\verb| http://unedf.org/content/highlights/tdslda-highlight-ascr_v3.pdf |

\bibitem{EN10}
S. Ebata, T. Nakatsukasa, T. Inakura, K. Yoshida, Y. Hashimoto and K. Yabana, 
Phys. Rev. C{\bf 82} (2010) 034306. 

\bibitem{BR86} 
J.-P. Blaizot and G. Ripka, 
Quantum Theory of Finite Systems (MIT Press, Cambridge, MA, 1986)

\bibitem{KB90}
S. J. Krieger, P. Bonche, H. Flocard, P. Quentin and M. S. Weiss, 
Nucl. Phys. {\bf A517} (1990) 275. 

\bibitem{CM80} 
R. Y. Cusson, J. A. Maruhn and H. St$\ddot{\rm o}$cker, 
Z. Physik. A{\bf 294} (1980) 257. 
\end{thebibliography}
\end{document}